\documentclass[authoryear,12pt,5p,times]{elsarticle}

\usepackage{bm}
\usepackage{xspace}
\renewcommand{\vec}[1]{\bm{#1}}
\newcommand{\matr}[1]{\bm{\mathsf{#1}}}

\newcommand{\loc}{\mathop{\mathrm{loc}}}
\newcommand{\expect}{\mathbb{E}}
\newcommand{\disp}{\mathbb{D}}
\newcommand{\FAP}{{\rm FAP}}

\usepackage{amssymb}
\usepackage{amsmath}

\journal{Astronomy and Computing}

\begin{document}
\sloppy

\begin{frontmatter}



\title{Detecting multiple periodicities in observational data with the multifrequency
periodogram -- II. Frequency Decomposer, a parallelized time-series analysis algorithm}


\author{Roman V. Baluev}

\address{Central Astronomical Observatory at Pulkovo of Russian Academy of Sciences,
Pulkovskoje sh. 65, St Petersburg 196140, Russia}
\address{Sobolev Astronomical Institute, St Petersburg State University, Universitetskij
pr. 28, Petrodvorets, St Petersburg 198504, Russia}
 \ead{roman@astro.spbu.ru}

\begin{abstract}
This is a parallelized algorithm performing a decomposition of a noisy time series into a
number of sinusoidal components. The algorithm analyses all suspicious periodicities that
can be revealed, including the ones that look like an alias or noise at a glance, but
later may prove to be a real variation. After selection of the initial candidates, the
algorithm performs a complete pass through all their possible combinations and computes
the rigorous multifrequency statistical significance for each such frequency tuple. The
largest combinations that still survived this thresholding procedure represent the outcome
of the analysis.

The parallel computing on a graphics processing unit (GPU) is implemented through CUDA and
brings a significant performance increase. It is still possible to run FREDEC solely on
CPU in the traditional single-threaded mode, when no suitable GPU device is available.

To verify the practical applicability of our algorithm, we apply it to an artificial time
series as well as to some real-life exoplanetary radial-velocity data. We demonstrate that
FREDEC can successfully reveal several known exoplanets. Moreover, it detected a new
$9.8$-day variation in the Lick data for the five-planet system of 55~Cnc. It might
indicate the existence of a small sixth planet in the 3:2 commensurability with the planet
55~Cnc~b, although this detection is model-dependent and still needs a detailed
verification.
\end{abstract}


\end{frontmatter}


\section{Introduction}
\label{sec_intro}
Hardly someone would object against the assertion that the extraction of a multiperiodic
variation in a raw time series data is one of the most importaint tasks of the practical
astronomy. Among the most relevant branches we may highlight, for instance, the
investigation of variable stars and the exoplanets searches. It is also widely known that
this task is often dramatically complicated by undesired but typical properties of the
data that are acquired by astronomers \citep{Vio13}. Such data are typically non-uniform;
moreover, they often demonstrate various regular, pseudo-regular, as well as irregular
gapping patterns that might get into severe interference with the real periodic
variations, which interfere between each other too. All this takes place above some
background noise, which has an a priori unknown (or only poorly known) variance. Since the
time when the \citet{Schuster1898} and the \citet{Lomb76}-\citet{Scargle82} periodograms
were introduced, a lot of efforts were done to overcome various issues arising in the task
of the spectral data analysis. These efforts were done in the field of theory work as well
as in the field of practical computing. We may highlight, in particular, that parallel
algorithms of periodogram computation using graphics processing units (GPUs) are getting
popularity in recent time \citep{Townsend10}.

Here we present a computation algorithm that may significantly facilitate this analysis.
It is meant to be a practical extension of our previous theory work \citep{Baluev13d},
hereafter Paper~I. In that work we provided an analytic approach to treat and compute the
multifrequency detection false alarm probabilities (hereafter $\FAP$). In particular, it
was demonstrated in Paper~I that to rigorously prove the simultaneous existence of
\emph{each} of $n$ presumably detected periodic components of a multiperiodic variation,
it is insufficient to just test each of the $n$ periodicities individually. It is
mandatory to additionally ensure that all these periodic components are statistically
significant \emph{jointly}, i.e. as a tuple. Also, it is necessary to verify that there is
enough statistical significance for each possible subtuple of any dimension $m<n$. Only
after all these statistical tests ($2^n-1$ tests in total) are passed through, we may
fairly claim that each of these $n$ periodicities likely exist (with a stated statistical
confidence, of course). Paper~I also contains an analytic approach to compute the false
alarm probabilities that are associated to the mentioned multifrequency tests. These
analytic approximations represent the multifrequency extensions of the ones that we
previously constructed for the single-frequency (e.g. Lomb-Scargle) periodograms
\citep{Baluev08a}.

Our computation algorithm, named as FREquency DEComposer (FREDEC), implements this theory
in a ready-for-use pipeline. The package can be dowloaded at
\texttt{http://sourceforge.net/projects/fredec/}. At first, it applies a consequent scan of
single-frequency periodograms to create an initial pool of candidate periodicities. This
preliminary scan represents some mixture of the QUICK and SLICK algorithms described by
\citet{Foster95}. Then each frequency combination of the constructed frequency pool is
considered in view of its complete multifrequency statistical significance. In the end,
the algorithm prints out the set of the largest independent frequency combinations that
were still found significant.

FREDEC is based on the multiperiodic model of an observable variation. This model
represents the sum of a limited number of sinusoids. Thus, it should perform well in the
cases when the actual variation can be well approximated by such a model, especially if
the exact model of the variation is unknown or too complicated. The suitable astronomical
cases include, for example, the exoplanetary signatures in stellar radial velocity
variations and variable stars of several types. This method is not suitable for e.g.
aperiodic variations (cathaclismic variables) or severely non-sinusoidal periodicities
(eclipsing binaries, exoplanetary transits, Doppler binaries/exoplanets involving orbital
eccentricities of $0.8$ or larger). In the latter case, we may need too large number of
sinusoidal harmonics to approximate the non-sinusoidal shape sufficiently well.

FREDEC is intended to run on a GPU device in a parallel regime, which increases its
performance dramatically. The GPU computing is implemented through the CUDA language. When
no suitable GPU device is available, the computations can be still done on CPU in a
conventional single-threaded manner.

The structure of the paper is as follows. In Section~\ref{sec_def}, we describe the main
definition and the analytic theory used by FREDEC. In Section~\ref{sec_locfit}, we
describe the core procedure of the algorithm~--- the non-linear fitting of the
multifrequency model. In Section~\ref{sec_pipeline} we provide a detailed description of
the entire algorithm pipeline. In Section~\ref{sec_GPU} we consider some GPU
parallelization issues of the algorithm. Finally, in Section~\ref{sec_treat} we give some
recommendations concerning the treatment of the FREDEC results. In Section~\ref{sec_app}
we discuss the application of our algorithm to several artificial as well as real-life
data-analysis examples.

\section{The definitions, the task layout, and the basic underlying theory}
\label{sec_def}
Let us have a time series containing of $N$ times $t_i$, measurements $x_i$, and weights
$w_i$. We will treat these data as the sum $x_i = \mu(t_i) + \epsilon_i$, where $\mu$ is a
parametric signal model that depends on the hypothesis adopted, and $\epsilon_i$ are
Gaussian and uncorrelated measurement errors. Concerning $\epsilon_i$, we will always
assume that $\expect \epsilon_i=0$ and $\disp\epsilon_i = \kappa/w_i$, where the common
multiplier $\kappa$ is unknown (it will be implicitly estimated from the data). We assume
that all frequencies that might exist in the data are located somewhere in a wide range
$[0,f_{\rm max}]$. The width of this frequency range is therefore equal to $f_{\rm max}$.
Using the effective time span $T_{\rm eff}=\sqrt{4\pi {\rm Var}(t_i)}$, where ${\rm
Var}(t_i)$ is the weighted variance of $t_i$, we can also define a non-dimensional
frequency bandwidth $W = f_{\rm max} T_{\rm eff}$, which plays an important role in
various false alarm probability estimations.

Our most basic null hypothesis involves the following data model:
\begin{equation}
\mathcal H_0 : \quad \mu(t) = c
\label{H0}
\end{equation}
where $c$ is an unknown constant to estimate. In fact, our algorithm may be also extended
to have a time polynomial in~(\ref{H0}) instead of just a constant $c$, but currently we
limit our attention to the case of only a free constant in $\mathcal H_0$.

We will deal below with multifrequency hypotheses that in general have the following
form:
\begin{eqnarray}
\mathcal H_n &:& \quad \mu(t) = c + \nonumber\\
&+& \sum_{k=1}^n a_k \cos(2\pi f_k t) + b_k \sin(2\pi f_k t).
\label{Hn}
\end{eqnarray}
Here, $c$, $a_k$ and $b_k$ are unknown linear coefficients, while the frequencies $f_k$
are unknown non-linear parameters. For each $\mathcal H_n$, all the parameters $c$, $a_k$,
$b_k$, and $f_k$, should be estimated from the data using the least-square regression.

Denote the averaging operator $\langle * \rangle$ as
\begin{equation}
\langle \phi(t) \rangle = \sum_{i=1}^N w_i \phi(t_i).
\end{equation}
and define the goodness-of-fit function, or the $\chi^2$ function, as
\begin{equation}
\chi^2_{\mathcal H_n}(\vec\theta,\vec f) = \left. \left\langle (x-\mu)^2 \right\rangle \right|_{\mathcal H_n},
\end{equation}
where the vector $\vec\theta$ contains all mentioned linear parameters, while the vector
$\vec f$ contains the frequencies. To solve the associated least-square regression task,
we must find the best-fit parametric estimates by means of minimizing the relevant
$\chi^2$ function. This can be split in two nested subtasks. The inner one involves only
an easy linear minimization
\begin{equation}
\vec\theta^*(\vec f) = \arg\min_{\vec\theta} \chi^2_{\mathcal H_n}(\vec\theta,\vec f),
\end{equation}
which can be performed extactly. In the outer subtask, we should perform a more difficult
non-linear fitting
\begin{equation}
\vec f^* = \arg\min_{\vec f} \chi^2_{\mathcal H_n}(\vec\theta^*(\vec f), \vec f), \quad
\vec\theta^{**} = \vec\theta^*(\vec f^*),
\end{equation}
which needs some iterative procedure. Below we will have rather little interest in the
best fitting parametric values $\vec\theta$ themselves. The quantities that will be more
important for us are the relevant minima of the $\chi^2$ function that eventually define
the signal significance. We denote them as
\begin{eqnarray}
l_n(\vec f) &=& \min_{\vec\theta} \chi^2_{\mathcal H_n} = \chi^2_{\mathcal H_n}(\vec\theta^*(\vec f),\vec f), \nonumber\\
l^*_n &=& \min_{\vec\theta,\vec f} \chi^2_{\mathcal H_n} = l_n(\vec f^*),
\end{eqnarray}

The multifrequency test statistic that measures how much $\mathcal H_n$ fits the data
better than $\mathcal H_0$, can be now written down as
\begin{eqnarray}
z_n(\vec f) &=& \frac{N_{\mathcal H_n}}{2} \log \frac{D-l_0}{D-l_n(\vec f)}, \nonumber\\
z^*_n &=& \frac{N_{\mathcal H_n}}{2} \log \frac{D-l_{\mathcal H_0}}{D-l^*_n(\vec f)} = \max_{\vec f} z(\vec f),
\label{zpow}
\end{eqnarray}
with $D=\langle x^2\rangle$, $N_{\mathcal H_n}=N-\dim \mathcal H_n=N-3n-1$, and $\dim \vec
f=n$. The first quantity defined in~(\ref{zpow}), $z(\vec f)$, is an intermediary one; it
formally corresponds to an assumption that all frequencies in $\vec f$ are known a priori,
and it only needs to solve a linear regression task. The second quantity, $z^*$,
corresponds to a general global test. These definitions take into account the unknown
noise scaling factor $\kappa$, which is implicitly reduced.

The formulae~(\ref{zpow}) represent a slight modification of the periodogram $z_3$ from
\citep{Baluev08a}. The frequency argument is now multidimensional, and the coefficient
$N_{\mathcal H_n}$ is reduced by the extra degrees of freedom introduced by the frequency
variables (in addition to the degrees of freedom provided by $\vec\theta$). The latter
modification is rather cosmetic. It does not change the asymptotic properties of the
periodogram (the relative difference decreases as $\sim 1/N$), which we will rely upon
below. This change in the coefficient was introduced mainly to make the algorithm more
conservative when dealing with small or moderate values of $N$.

In addition to the global test~(\ref{zpow}), we define the local multifrequency test,
which is computationally much faster. Let us have some approximate preliminary frequencies
estimation in the vector $\vec f_{\rm loc}$. These preliminary frequencies typically
represent the positions of some periodogram peaks. We assume that the true frequencies are
indeed located inside of these peaks; they only need to be locally refined using the
complete multifrequency model. In this case we can treat the model~(\ref{Hn})
well-linearizable with respect to $f_i$, so we can apply some gradient method of
non-linear minimization, starting from the initial position of $\vec f_{\rm loc}$. What we
get in the end of the iterations is the nearest local minimum $l_{n,\rm loc}^*$ and the
implied local test statistic $z^*_{n,\rm loc}$. Hereafter we will denote such local maxima
near $\vec f_{\rm loc}$ as
\begin{equation}
z^*_{n,\rm loc}(\vec f_{\rm loc}) = \loc\max_{\vec f \approx \vec f_{\rm loc}} z_n(\vec f)
\end{equation}
Clearly, this $z^*_{\rm loc}$ is a discontinuous function: when some frequency in $\vec
f_{\rm loc}$ passes between neighbouring periodogram peaks, the value of $z_{\rm loc}$
changes abruptly at some boundary point. To compute the global maximum $z^*$, we need to
sample $z^*_{\rm loc}$ over a dense enough multidimensional grid (considering that the
natural frequency resolution is $1/T$), and then to find the maximum.\footnote{Notice that
this frequency grid may be more rarified than the one that we would need to use when
determining $z^*$ by a ``brute force'' maximization of $z(\vec f)$.}

We call the test in~(\ref{zpow}) as \emph{absolute}, because it provides an absolute
likelihood of the best-fit $n$-frequency tuple. Eventually, we will need \emph{relative}
tests that compare two nested frequency tuples with each other. The relevant
fixed-frequency test statistic (analogue of $z_n(\vec f)$) can be defined as
\begin{equation}
z_{n|m}(\vec f|\vec f') = \frac{N_{\mathcal H_{n+m}}}{2} \log \frac{D-l_m(\vec f')}{D-l_{n+m}(\vec f',\vec f)}.
\label{zpowrel}
\end{equation}
Here $\vec f'$ is an $m$-frequency tuple that corresponds to the base model $\mathcal
H_m$. The alternative model $\mathcal H_{n+m}$ involves $m$ base frequencies $\vec f'$ and
also an additional set of $n$ frequencies $\vec f$. This relative test statistic defines
the likelihood of $n$ given frequency components under the assumption that $m$ other
frequencies are already established. It is also assumed that all related frequency values
are known precisely and thus are fixed.

To derive from~(\ref{zpowrel}) a variable-frequency case, we must recall that the base
model $\mathcal H_m$ is useful only when it is understood in the local sense. We assume
that there exist $m$ approximately-known frequencies $\vec f'$: they are allowed to vary
within a narrow neighborhood of $\vec f_{\rm loc}'$. Given this base model, how realistic
would be an expanded model with $n$ extra frequencies $\vec f$? When $\vec f$ is still
fixed, the relevant likelihood-ratio measure may be defined with the formulae:
\begin{eqnarray}
z^*_{n|m,\rm loc}(\vec f|\vec f'_{\rm loc}) &=& \frac{N_{\mathcal H_{n+m}}}{2}
 \log \frac{D-l_{m,\rm loc}(\vec f_{\rm loc}')}{D-l_{n|m,\rm loc}(\vec f|\vec f_{\rm loc}')}, \nonumber\\
l_{m,\rm loc}(\vec f_{\rm loc}') &=& \loc\max_{\vec f'\approx \vec f_{\rm loc}'} l_m(\vec f'), \nonumber\\
l_{n|m,\rm loc}(\vec f|\vec f_{\rm loc}') &=& \loc\max_{\vec f'\approx \vec f_{\rm loc}'} l_{n+m}(\{\vec f',\vec f\}),
\end{eqnarray}
Optimizing out the variable $\vec f$ too, we introduce the following double-local and
global-local tests:
\begin{eqnarray}
z^*_{n,\rm loc|m,\rm loc}(\vec f_{\rm loc}|\vec f'_{\rm loc}) &=&
 \loc\max_{\vec f \approx \vec f_{\rm loc}} z^*_{n|m,\rm loc}(\vec f|\vec f'_{\rm loc}), \nonumber\\
z^*_{n|m,\rm loc}(*|\vec f'_{\rm loc}) &=&
 \max_{\vec f} z^*_{n|m,\rm loc}(\vec f|\vec f'_{\rm loc}).
\end{eqnarray}

Let us assume that we have detected $n$ possible periodic components exist in the data;
these components are defined by a preliminary frequency vector $\vec f_{\rm loc}$. As we
discuss in Paper~I, to verify that all of these components are indeed statistically
significant, we must apply $2^n-1$ statistical tests in total. These are the relative
tests $z^*_{n-m,\loc|m,\loc}(\vec f_{\rm loc}|\vec f_{\rm loc}')$, where $\vec f_{\rm
loc}'$ is an arbitrary $m$-dimensional subvector of $\vec f_{\rm loc}$. For each integer
$m$ from $0$ to $n-1$ we have $C_n^m$ of such multifrequency tests, so their total number
counts to $2^n-1$.

Even though all the putative components have passed \emph{individual} single-frequency
tests, this does not guarantee that all their combinations will pass the \emph{joint}
multifrequency tests too. If just a single such combination yields insufficient
significance then we have to admit that some of the frequencies in $\vec f_{\rm loc}$
still may be fake: they may prove as a noise artifact or an alias.

For example, when two frequencies are individually significant but do not score enough
joint significance, this means that we cannot claim that both these components are
``detected'', even if these components generate equal peaks on the periodogram and are not
mutual aliases. In this case we should just select these two single-frequency components
as peer explanations of the data, without combining them together. What we can say for
sure is that at least one of these periodicities likely exists. Whatever periodicity we
adopt as true, either this or another one might be confirmed as well as disproved later.
We have insufficient observational basis to simultaneously select them both, but we cannot
reject them both as well.

The multifrequency test statistics that we have defined above are not calibrated yet.
Under ``calibration'' of a test statistic $z$ we mean basically a mapping that can
transform each $z$-value to the associated false alarm probability, $\FAP(z)$. Note that
because we did not knew the vector $\vec f_{\rm loc}$ in advance, the $\FAP$ must be
calculated as if we have run a full scan of the frequency space, i.e. as if we used the
global-local statistic $z^*_{n-m|m,\loc}(*|\vec f_{\rm loc}')$ everywhere, even though we
might actually compute its double-local version $z^*_{n-m,\loc|m,\loc}(\vec f_{\rm
loc}|\vec f_{\rm loc}')$. The latter statistic is used just as a rapid computational, but
not analytic, replacer for the former one after $\vec f_{\rm loc}$ is obtained.

Now we need to adapt the main results of Paper~I, where we have constructed the $\FAP$
estimations for some multifrequency test statistics. Those results are still not matching
our needs perfectly. First, they refer to only absolute tests similar to $z^*_n$
in~(\ref{zpow}) rather than to $z^*_{n|m,\rm loc}$. Secondly, this $\FAP$ approximation
refers only to a simplified version of $z^*_n(\vec f)$, corresponding to the case when the
uncertainties of $\epsilon_i$ are known exactly (rather than expressed through $w_i$).
However, as we have discussed in Paper~I, these simplified $\FAP$ expressions still can be
used as asymptotic ($N\to\infty$) approximations to the $\FAP$ for the periodograms that
we denoted here as $z^*_{n|m,\rm loc}$. This is because the base multifrequency models of
these statistics, as well as the multiplicative noise model, are understood in the local
sense. The relevant non-linear parameters (the frequencies and $\kappa$) thus appear
well-linearizable.

Therefore, the FREDEC code relies on the following multifrequency $\FAP$ formula from
Paper I:
\begin{equation}
\FAP_n(z) \lesssim M_n(z) \simeq \tilde A_n W^n e^{-z} z^{\frac{3n}{2}-1},
\label{zFAP}
\end{equation}
where $\tilde A_n$ are some numeric coefficients that we do not detail here. We use this
formula for all multifrequency periodograms of the type $z^*_{n|m,\loc}(*|\vec f_{\rm
loc}')$, and consequently for their computational replacers $z^*_{n,\loc|m,\loc}(\vec
f_{\rm loc}|\vec f_{\rm loc}')$. Obviously, the formula~(\ref{zFAP}) is invariable with
respect to $\vec f_{\rm loc}$; i.e. the periodogram's detection levels in the first
approximation do not depend on the parameters of the base model (although the periodograms
themselves do depend on them, of course).

In general, the FREDEC algorithm is doing the following: (i) it constructs a wide enough
initial pool of $n$ preliminary frequencies in the vector $\vec f_{\rm loc}$; (ii) it
computes the set of all necessary test statistics $z^*_{m-k,\loc|k,\loc}$; (iii) it tests
each independent multifrequency combination (a subvector of $\vec f_{\rm loc}$), keeping
only the largest combinations that still pass the multifrequency $\FAP$ threshold based
on~(\ref{zFAP}).

How many tests we should apply during this sequence? We can sample $C_n^m$ independent
$m$-frequency combinations $\vec f_{\rm loc}'$ out of the original $n$-frequency pool
$\vec f_{\rm loc}$. For each such combination we must compute $2^m-1$ relative test
statistics to ensure its statistical significance. In each of these statistics,
$z^*_{m-k,\loc|k,\loc}$, the combination $\vec f_{\rm loc}'$ is split in two subsets
having sizes of $m-k$ and $k$ (for $0\leq k\leq m-1$) that serve as the arguments of the
statistic. For a given $m$ and $k$ the number of such statistics is $C_m^k$ (clearly, they
sum to $2^m-1$, as expected). Therefore, the total number of the tests to apply to the
original pool is equal to $\sum_{m=1}^n C_n^m (2^m-1) = 3^n - 2^n$. This is a quickly
growing function that will inevitably limit us to only rather moderate numbers $n$. Of
course, this algorithm still can be optimized in several directions, which are discussed
below.

\section{Computing the local multifrequency fit}
\label{sec_locfit}
The core procedure of the FREDEC algorithm is the computation of the local $\chi^2$ minima
that we have denoted as $l_n$ and $l_{n,\rm loc}$. The first function requires to carry
out a linear least-square minimization:
\begin{equation}
\chi^2_{\mathcal H_n}(\vec\theta,\vec f) = D - \vec g(\vec f) \cdot \vec\theta +
 \frac{1}{2} \vec\theta^{\rm T} \matr Q(\vec f) \vec\theta \longmapsto \min_{\vec\theta}.
\label{lsq}
\end{equation}
Here we have represented the $\chi^2$ function through a quadratic form, which is possible
thanks to the linearity of $\vec\theta$. The likelihood function gradient $\vec g$ and the
Fisher matrix $\matr Q$ both are functions of the frequencies. They can be expressed as
\begin{eqnarray}
\vec g &=& \Big\{ \langle x \rangle, \langle x \cos \omega_1 t \rangle, \langle x \sin \omega_1 t \rangle, \nonumber\\
  & & \langle x \cos \omega_2 t \rangle, \langle x \sin \omega_2 t \rangle, \ldots, \nonumber\\
  & & \langle x \cos \omega_n t \rangle, \langle x \sin \omega_n t \rangle \Big\}
\label{grad}
\end{eqnarray}
and
\begin{equation}
\matr Q = \left(\begin{array}{@{}c@{\,}c@{\,}c@{\,}c@{}}
  \langle 1 \rangle                               & \langle \cos\omega_1 t \rangle & \langle \sin\omega_1 t \rangle & \ldots \\
  \langle \cos\omega_1 t \rangle & \langle \cos^2 \omega_1 t \rangle & \langle \sin\omega_1 t \cos\omega_1 t \rangle & \ldots \\
  \langle \sin\omega_1 t \rangle & \langle \sin\omega_1 t \cos\omega_1 t \rangle & \langle \sin^2 \omega_1 t \rangle & \ldots \\
  \ldots & \ldots & \ldots & \ldots
 \end{array}\right),
\label{fisher}
\end{equation}
where $\omega_k = 2\pi f_k$, and the dots stand for the elements containing other
$\omega_k$ analogously to the shown ones with $\omega_1$. The general definition of $\vec
g$ and $\matr Q$ can be found in Paper~I.

The solution to the task~(\ref{lsq}) is explicit: $l_n = D - \vec g^{\rm T} \matr Q^{-1}
\vec g /2$. A quick way to compute $l_n$ is to apply the Cholesky decomposition $\matr
Q=\matr L \matr L^{\rm T}$, where $\matr L$ is a low-triangular matrix. Then we can
compute $\vec a=\matr L^{-1}\vec g$ using a forward substitution of $\vec g$, and finally
we have $D-l_n=\vec a^2/2$. The associated best fitting parameters can be expressed as
$\vec\theta^* = (\matr L^{\rm T})^{-1} \vec a$, which can be computed by a back
substitution of $\vec a$.

Fitting of the frequencies $\vec f$ is an iterative non-linear procedure, which involves
the fitting of $\vec\theta$ as a subtask. Assume that we have already performed the linear
fit of $\vec\theta$ and need to refine $\vec f$ and $\vec\theta^*$. Now we can write down
the following quadratic approximation:
\begin{equation}
\chi^2_{\mathcal H_n}(\vec\theta,\vec f) = D - \vec g_{\vec f} \cdot \Delta\vec\xi +
 \frac{1}{2} \Delta\vec\xi^{\rm T} \matr Q_{\vec f} \Delta\vec\xi + \ldots,
\label{lsqf}
\end{equation}
where the vector $\Delta\vec\xi$ encapsulates the parametric steps $\Delta \vec\theta$ and
$\Delta\vec f$. The vector $\vec g_{\vec f}$ is the likelihood function gradient over
$\vec\xi$. It is similar to $\vec g$, but must be computed for
$\vec\theta=\vec\theta^*(\vec f)$, where $\vec f$ is the frequency vector of the current
iteration. The first part of $\vec g_{\vec f}$, which is associated to the parameters
$\vec\theta$, is necessarily zero, because it was annihilated during the linear fitting
stage. The low-top submatrix of $\matr Q_{\vec f}$ coincides with $\matr Q$. The non-zero
subvector of $\vec g_{\vec f}$ and the remaining parts of $\matr Q_{\vec f}$ depend on the
values of $\vec\theta^*$ that were obtained previously. These elements involve, in
particular, the averaged derivatives of the model~(\ref{Hn}) over the frequency vector
$\vec f$.

Since $\matr Q$ is a low-top submatrix of $\matr Q_{\vec f}$, the Cholesky matrix $\matr
L$ is also a low-top submatrix of $\matr L_{\vec f}$ (the Cholesky matrix for $\matr
Q_{\vec f}$). Therefore, we do not need to apply the Cholesky decomposition anew. It can
be easily implemented in an incremental manner, extending the pre-calculated $\matr L$ to
$\matr L_{\vec f}$. After the Cholesky decomposition is completed, we can compute the
implied parametric step $\Delta\vec\xi = \matr Q_{\vec f}^{-1} \vec g_{\vec f}$, refine
the frequency vector, and proceed to the next iteration, which will start from the linear
fitting again. After we reach a satisfactory accuracy in $\vec f$, we still need to run
the linear fitting subroutine once again to compute $l_{n,\rm loc}$, which we originally
aimed to obtain.

We would like to highlight that the fitting algorithm that we presented above is more
efficient than a general non-linear fitting algorithm. We significantly profit here from
the linearity of the parameters $\vec\theta$, which allows for more accurate iterations.
The iteartions are more accurate because instead of using the values of $\vec\theta$ from
a previous iteration, we first refine them to honour the latest update of $\vec f$. Thanks
to re-using of the matrix $\matr Q$, no significant overheads are implied. This approach
is generally similar to the one suggested by \citet{WrightHoward09} for exoplanetary fits
of radial velocity data.

\section{The FREDEC pipeline}
\label{sec_pipeline}
\subsection{Initialization}
In addition to some variables initialization, data loading, and GPU hardware
initialization, we perform some useful normalizations of the time series. These
normalizations are intended to fulfil the following relations:
\begin{eqnarray}
\langle 1 \rangle = 1, \quad \langle x \rangle = 0, \quad \langle x^2 \rangle=1,\nonumber\\
\langle t \rangle = 0, \quad \langle t^2 \rangle = 1.
\label{idn}
\end{eqnarray}
These relations are very useful to satisfy, because they considerably simplify the
computation formulae for the elements in~(\ref{grad}) and~(\ref{fisher}) and for some
other similar quantities. Otherwise, we would have to carry or re-evaluate the quantities
in the left hand sides of~(\ref{idn}) through all algorithm pipeline. For example, these
relations imply the identity $\langle\cos^2 \omega t\rangle + \langle\sin^2 \omega
t\rangle = 1$, which allows us to omit the evaluation of some of the elements in the
matrix $\matr Q$.

\subsection{Phase 1: preliminary scan}
During this phase we must create the basic pool of candidate frequencies. The most honest
and direct way to do so is to run a full multidimensional scan of an $n$-frequency
periodogram with some large enough $n$. However, this is obviously not practically
feasible, so we need to apply some other method. We use a mixture of the QUICK and SLICK
algorithms described by \citet{Foster95}. We compute a series of the single-frequency
residual periodograms, each time adding to the base model the frequency corresponding to
the largest peak remaining. This is the SLICK part of the scan. The final pool of the
candidate is not limited, however, by the highest peaks of each of these sequential
periodograms. We also honour other periodigram peaks that demonstrated small enough
single-frequency $\FAP$. These side peaks do not go to the set of the base frequencies to
be used when constructing the next residual periodogram, but they go to the final pool of
the candidates. This is the QUICK part of the scan. In such a way, our final pool will be
probably overfilled, i.e. it will likely contain some aliases or even noisy peaks. We
avoid to do any conclusions at this early stage, however, because the peak that initially
looked as an alias may later appear as true. On contrary, real variations may initially
look as false peaks sometimes \citep{Foster95}.

The comprehensive set of the conditions that a periodogram peak must satisfy to go to the
pool is:
\begin{enumerate}
\item Its single-frequency $\FAP$, calculated from~(\ref{zFAP}) substituting $n=1$ is
smaller than some settled threshold $\FAP_1$. The $\FAP_1$ threshold might be rather mild
(we use $0.1$ by default).

\item Its height is at least half of that of the maximum peak found on this periodogram.
This condition is a workaround to handle the situation when the data contain a single
dominating variation, which generates a lot of large alias peaks obscuring smaller
variations that would reveal themselves after removal of the dominating one.
\end{enumerate}

The subsequent residual periodograms are computed until the maximum peak's $\FAP$ rises
above another threshold $\FAP_0$. Obviously, the inequality $\FAP_1 \geq \FAP_0$ must be
satisfied for the algorithm to be logically self-consistent.

Sometimes the candidates pool may grow too much. To prevent this, we set an upper limit of
$N/10$ on its size. Candidates with the largest detection $\FAP$ that are out of this
limit by the end of Phase~1 are just thrown away. Since each periodicity requires three
parameters in the model~(\ref{Hn}), the largest ever possible number of the free
parameters is thus equal to $\sim N/3$.

\subsection{Phase 2: forward cascade pass}
During this phase, the algorithm computes the set of the values of $l_{m,\rm loc}$ for all
possible subsets drawn from the pool of the candidates in all possible combinations. There
are $C_n^m$ independent absolute $m$-frequency tests for each $m=1,2,\ldots,n$. Usually
this computation stage is the heviest one. The number of the values to compute is $2^n-1$.

\subsection{Phase 3: backward cascade pass}
Based on the previously calculated values of $l_{m,\rm loc}$, we can now comute the values
of all necessary relative test statistics $z^*_{m-k,\loc|k,\loc}$, for $k=1,2,\ldots m$,
and then to apply the $\FAP$ threshold to them. This phase does not require any non-linear
minimization or the expensive averaging of the trigonometric functions, like the phase~2,
but the number of the quantities to compute is now increased to $\sim 3^n$. Without extra
optimizations, this apparently insignificant change makes the phase~3 computation to run
even slower than the phase~2, when $n$ exceeds $\sim 20-25$.

First, we can avoid the computation of the $\FAP$, which involves transcendent functions,
for \emph{each} test statistic. Instead, we may find the minimum (i.e., the worst-case
value) among all $z^*_{m-k,\loc|k,\loc}$ belonging to a layer with the same $k$, and only
after that we should pass this minimum to the $\FAP$ threshold. This is because $\FAP$ for
the same $k$ is expressed by the same formula. However, the layers of the tests with
different $k$ may be only compared in terms of the $\FAP$, because the
formula~(\ref{zFAP}) depends on the dimensionality of the model.

Secondly, we do not actually need to \emph{compute} $\FAP$s, we need to \emph{threshold}
them. For some $m$-frequency combination $\vec f'$, sampled out of the original
$n$-frequency pool $\vec f$, there are $2^m-1$ relative tests to compute, each referring
to some lesser subsample of $\vec f'$. But this computation can be interrupted right after
we found a subsample that failed the significance test. In case of such a fail we can
immediately proceed to the next combination $\vec f'$, skipping any further subsamples
from the current $\vec f'$. The complete $\FAP$ of the frequency tuple is the maximum
among the $\FAP$s of the subsampled combinations, and once this maximum exceeded the
threshold, it will never return below it. To further increase the performance, we may
alternate the values of $k$ so that the largest test layers (with $k\sim m/2$) are left
for later; this will increase the chance that some test will fail before we get to the
most complicated part of the job.

With these optimizations, the phase~3 computation time was dramatically reduced, and even
became negligible in comparison with the phase~2.

The $\FAP$ thresholding during the phase~3 is controlled by an additional parameter
$\FAP_2$, and it should not exceed $\FAP_0$ or $\FAP_1$ to preserve the logical
consistency of the algorithm. Therefore, the double inequality $\FAP_1 \geq \FAP_0 \geq
\FAP_2$ must be satisfied. Default values are: $\FAP_1=0.1$ and $\FAP_0=\FAP_2=0.05$.

\subsection{Phase 4: alternatives filtering}
The frequency combinations that survived the phase~3 form the output pool of alternative
multiperiodic models of the data. This does not imply, however, that all these
alternatives are statistically equivalent. In fact, the results of the algorithm often
contain frequency combinations that offer clearly bad fit of the data (in comparison with
the other ones). The only thing that is guaranteed is that the results will never contain
\emph{nested} frequency combinations.

To say that our work is completed we must carry out a statistical comparison between the
remaining \emph{non-nested} models. Testing of non-nested hypotheses is significantly
different from the more traditional nested hypotheses case \citep{Baluev12}. For the case
of only two rival hypotheses we could apply e.g. the Vuong test for this goal
\citep{Vuong89,Baluev12}. However, our case involves multiple alternative models, which
disables the direct use of the Vuong test. The case of the multiple non-nested hypotheses
still needs some more deep theoretic investigation.

Therefore, this phase~4 of the FREDEC pipeline is currently incomplete. The present
version of FREDEC only sorts out the alternatives in the $\chi^2$-increase order to make
it easier for the user at least to identify the models that offer a clearly bad fit. Also,
the algorithm computes the set of values of the Vuong statistic comparing the best fit
with all others. Since the application of this test to multiple alternative hypotheses is
not currently very rigorous, these values should be treated with care. Nevertheless,
FREDEC allows to filter out only the alternatives that have the Vuong statistic smaller
than some critical value. We set this threshold to a rather conservative level of $5$ by
default.

\section{GPU parallelization}
\label{sec_GPU}
Profiling tools show that more than $90\%$ of the FREDEC computing time is spent during
the evaluation of the sine and cosine functions. Actually, the same proposition is true
for the classic Lomb-Scargle periodogram. Therefore, the most of the computing resources
are spent for the trigonometric averages that appear in the gradient vector $\vec g$ and
matrix $\matr Q$, as well as in their extensions $\vec g_{\vec f}$ and $\matr Q_{\vec f}$.
These averages can be split in two independent systems. The first system is used to
evaluate the gradient:
\begin{eqnarray}
\langle \cos\omega t \rangle, \quad \langle \sin\omega t \rangle, \nonumber\\
\langle t \cos\omega t \rangle, \quad \langle t \sin\omega t \rangle, \nonumber\\
\langle x \cos\omega t \rangle, \quad \langle x \sin\omega t \rangle, \nonumber\\
\langle x t \cos\omega t \rangle, \quad \langle x t \sin\omega t \rangle,
\label{avrg1}
\end{eqnarray}
where $\omega$ is equal to one of $\omega_k$. The second one is used to compute the
elements of the Fisher matrix:
\begin{eqnarray}
\langle \cos\omega t \rangle, \quad \langle \sin\omega t \rangle, \nonumber\\
\langle t \cos\omega t \rangle, \quad \langle t \sin\omega t \rangle, \nonumber\\
\langle t^2 \cos\omega t \rangle, \quad \langle t^2 \sin\omega t \rangle,
\label{avrg2}
\end{eqnarray}
where $\omega=\omega_k\pm\omega_m$, excluding the difference for $k=m$. The averages
involving the $t$ or $t^2$ multipliers are necessary to calculate $\vec g_{\vec f}$ and
$\matr Q_{\vec f}$; they appear due to the derivatives of~(\ref{Hn}) over $\vec f$.

The computation of~(\ref{avrg1}) and~(\ref{avrg2}) can be very efficiently parallelized on
GPU, since we need to evaluate the quantities of the same type differing only in the value
of $\omega$. Besides, all of these averages are based on the same time series data
$(t_i,x_i,w_i)$ that can be pre-loaded into the fast shared memory of the GPU. The
algorithm is generally similar to the one proposed by \citet{Townsend10} for the classic
Lomb-Scargle periodogram. The performance increase factor for this part of the FREDEC
algorithm is relatively high. It reaches hundreds on the top-class GPU (tested with NVIDIA
Tesla C2075), though it was smaller for less powerful GPU cards (we tested NVIDIA GeForce
210). This performance increase also significantly depends on the adopted floating-point
arithmetics~--- single- or double-precision. We however do not recommend to use single
precision for practical calculations with FREDEC due to large round-off errors leading to
numerical instability.

Most other parts of the algorithm are also adapted for GPU computing, although it seems
that their parallelization is not that efficient, maybe because of less efficient memory
usage. In particular, the parallel least-square fitting of Sect.~\ref{sec_locfit} is
implemented by means of launching of many entirely independent instances of the fitting
subroutine. However, the internal data arrays used of these fitters are all different and
have to be stored in a rather slow global GPU memory.

The overall performance increase with the mentioned NVIDIA Tesla GPU was $\sim 30$ for
double-precision arithmetics and $\sim 150$ for single precision. The difference between
the single- and double-precision tests was mainly due to a mysterious slow-down of the CPU
computation on single-precision, while the GPU benchmark demonstrated, on contrary, a
moderate speed-up. The mentioned NVIDIA GeForce card only supports single-precision
arithmetics, and in this case the GPU/CPU performance increase factor was $\sim 20$.

The performance of the algorithm depends severely on the number of the frequencies in the
initial pool, $n$. When this $n$ is smaller than $15$ the computation passes through
pretty quickly both in GPU and CPU mode. For $n=15-20$ the CPU computation will be long
though still feasible, while the GPU one is still rather fast. The values $n=25-27$
represent the limit of the FREDEC capabilities. In some pracical data that we considered
during the testing (they are the public radial velocity data for some exoplanet-hosting
stars), the maximum value of $n$ that we dealt with was $25$ (that was the case of the
Lick data for 55~Cancri, considered below), while other cases usually implied a
significantly smaller $n$.

\section{Interpretation of the FREDEC results}
\label{sec_treat}
The FREDEC output is a set of alternative multifrequency models. The computation pipeline
described above verifies that within each such model \emph{all} its periodic components
likely exist (at the significance level of $\FAP_2$). Presently, FREDEC does not provide a
unique and rigorous way to define which of these alternative models are likely and which
are not. As we have explained above, we need a more intricate method of multiple
non-nested hypotheses testing to do this part of the work. The output contains the
following data per each multiperiodic solution:
\begin{enumerate}
\item Best fitting frequency values $f_i$, sorted in the increase order.
\item The adimensional goodness-of-fit value $G=l_{m,\rm loc} N_{\mathcal H_0}/N_{\mathcal
H_m}$. Due to the normalization~(\ref{idn}), this quantity is equal to the ratio of the
reduced $\chi^2$ values for the best fits of the associated model $\mathcal H_m$ and of
the null model $\mathcal H_0$. The reduced $\chi^2$ value for $\mathcal H_0$ is the
classic variance estimation of the original (unscaled) $x_i$, taken with weights $w_i$.
Since this variance is the same over all the alternative fits, the quantity $G$ represents
just a scaled value of the reduced $\chi^2$ of the multiperiodic model. Smaller values of
$G$ correspond to more preferrable solutions, although we do not define any formal
probabilistic measure of the relevant advantage.
\item The Vuong statistic comparing this fit with the one offering the smallest value of
$G$. For large $N$, each individual Vuong statistic asymptotically follows a standard
normal distribution. However, since here we typically have more than two alternative
solutions, we have more than a single such comparison test, and when we apply many similar
tests, we get an increased chance to make a mistake. This effect of multiple hypothesis
testing should increase the thresholding level for the Vuong test, in comparison with the
quantile levels of the standard normal distribution. Thus the values of the Vuong test
reported by FREDEC are currently not calibrated well.
\item The single-frequency $\FAP$ associated to the maximum peak still remaining in the
residual periodogram. Small value of this $\FAP$ indicates that after subtraction of this
particular multiperiodic solution some significant periodic variations still remain in
the data. This may mean that either this solution is parasitic and should be rejected in
favour of another one or it is the correct one, but the data still contain some
significant residual variation that cannot be reliably decomposed.
\end{enumerate}
The values of $G$, of the Vuong statistic, and of the residual single-frequency $\FAP$ may
be used to filter out the solutions that provide clearly bad fit to the data. To be more
helpful here, FREDEC sorts the solution in the $\chi^2$-increase order (grouping them in
bunches with the same $m$). However, these criteria are currently unrigorous and indirect.
For example, it is rather normal when \emph{all} of the proposed solutions have small
residual $\FAP$, and even all below the $\FAP_2$ threshold.

Notice that we assumed a strictly multiperiodic model~(\ref{Hn}), and a strict
multiplicative model of the noise. In the case when either of the model might be
inaccurate, the results reported by the FREDEC are suggestive rather then decisive.

It is also important to pay attention to the construction of the initial pool of
candidates during the Phase~1. When FREDEC truncates this pool by a significant amount (to
keep its size below the limit of $N/10$), this indicates that the data set is to small to
provide a complete solution. In this case the data likely contain many periods, but it is
impossible to properly process all of them due to a large number of free parameters to
fit.

\section{Practical examples}
\label{sec_app}
\subsection{Double-frequency example from Paper~I}
In Paper~I we considered an artificial time series, containing two sinusoids at the
frequencies of $0.9$~Hz and $1.1$~Hz, and periodic data gaps generating an aliasing
frequency of $0.1$~Hz. The single-frequency periodogram of these data shows the maximum
peak at a wrong frequency of $1.0$~Hz, while the true frequencies look like some side
aliases. These data generate a sequence of detectable periods at the frequencies of
$(1.0\pm 0.1 k)$~Hz.

When applied to the original time series of Paper~I, our FREDEC algorithm correctly
identifies the double-frequency combination used to construct the data. However that data
set was entirely noiseless. It is more interesting to consider noisy data, so we added to
the original time series a small Gaussian noise with the standard deviation equal to
$1/10$ of the amplitudes of the original sinusoidal variations.

First of all, FREDEC again successfully identifies a single double-frequency solution with
the correct frequencies of $0.9$~Hz and $1.1$~Hz. This model has the value of $G$ close to
the minimum, and the Vuong statistic of $0.5$, indicating a pretty good fit. Additionally,
there are $14$ alternative combinations containing $5-7$ components involving various
aliased periods. Most of these models could be rejected due to a large value of the Vuong
statistic (up to $6.6$). The most likely combinations are: two solutions with $7$
components (one of the true frequencies and $6$ aliases in the range from $0.6$~Hz to
$1.4$~Hz), a single solution with $5$ components (aliases from $0.7$ to $1.3$~Hz without
the true frequencies), and the correct double-frequency solution.

These results indicate that the maximum periodogram peak at $1.0$~Hz may only lead us to
very complicated models containing no less than $5$ periodicities. The simplest admissible
model contains two frequencies that initially looked like mere aliases.

\subsection{Radial velocity data for the 51~Peg exoplanetary system}
We use the public ELODIE \citep{Naef04} radial velocity data for this famous
planet-hosting star. In the ELODIE data, FREDEC easily identifies the primary (planetary)
variation with the period of $4.2308$~d. However, a weak though clearly detectable
($\FAP\sim 10^{-9}$) additional variation is also revealed. Its period is subject to alias
ambiguity, and could be one of: $359.3$~d (the best fit), $23^h 52^m$ (Vuong statistic of
$0.8$), $24^h 00^m$ (Vuong statistic of $1.7$), and $24^h 04^m$ (Vuong statistic of
$2.7$). All these values are mutual aliases that likely reflect the presence of a
systematic annual variation in the ELODIE data. We have already detected this variation in
these data in our old work \citep{Baluev08b} by means of the traditional periodogram. Now,
FREDEC confirms this result and gives more details. No more periods in the ELODIE data are
seen.

\subsection{Radial velocity data for the GJ~876 exoplanetary system}
This planetary system is famous thanks to a detectable secular apsidal drift of the two
main planets \citep{Rivera10,Correia10}. In the radial velocity periodograms an apsidal
drift of a planet with an orbital period $P$ appears as a small shift of all related
overtone periods $P/k$. The unperturbed multi-Keplerian model of the radial velocity curve
cannot take this effect into account, but the multiperiodic model with freely fittable
frequencies can. In fact, we may expect that a mutiperiodic model may fit such data at an
accuracy level comparable to that of the rigorous Newtonian $N$-body model.

We run FREDEC separately for the HARPS \citep{Correia10} and Keck \citep{Rivera10} radial
velocity data. In the HARPS data we only robustly detect the periods of the two main
planets $P_b \approx 60$~d and $P_c \approx 30$~d. There was also the third ambiguous
period of either $\sim P_c/2$ or $\sim P_c/3$. We actually know that these overtone
periods exist \emph{simultaneously}, but FREDEC finds that their joint significance in the
HARPS data is too low, and suggests them as peer alternatives. In the output of Phase~1 of
the algorithm we also find a set of periods close to the period of the third planet
$P_d\sim 2$~d, but they were excluded from the analysis to comply with the maximum allowed
number of the components. This is not very surprising, since the number of the HARPS data
is still rather small to work entirely alone.

In the Keck data, the best FREDEC solution contains $6$ components, which involve all four
known planets of the system ($P_b \approx 61$~d, $P_c\approx 30$~d, $P_d\approx 2$~d, and
$P_e\approx 125$~d), and two subharmonics $\sim P_c/2$ and $\sim P_c/3$. Additionally to
this nominal solution, there are $35$ alternative models that involve various aliases
(typically the diurnal ones). Most of them can be rejected using the Vuong test: we find
only $6$ models having the Vuong statistic below $3$, all with $6$ components. Only one of
these remaining alternatives appears relatively non-trivial. It contains no period $P_e$,
and in place of the $P_c/2$ subharmonic it contains \emph{two} close periods of $15.0$~d
and $14.3$~d. We believe this reflects some effect of secular motion due to Newtonian
perturbations.

It must be noted that the RV data for GJ876 are affected by non-white noise
\citep{Baluev11}, which formally invalidates all statistical methods that FREDEC relies
on. However, in the FREDEC results described above we did not find any clear signature of
the correlated noise. Probably, in this case the correlated noise is partly obscured by
inaccuracies of the multiperiodic models.

\subsection{Radial velocity data for the 55~Cnc exoplanetary system}
This planetary system contains five known planets \citep{Fisher08}. Their orbital
eccentricities are small, as well as their gravitational perturbations. Therefore, the
multi-sinusoidal model should work well for these data. Application of the FREDEC
algorithm to the published Lick data for this star reveals dozens of alternative
solutions. However, most of them, even if pass the Vuong test, are not very likely because
they contain various periods close to one day. These periods appear due to diurnal
aliasing cycles of the data. Anyway all periods close to 1~day are unlikely, so we paid
attention to only non-diurnal periods. These periods are: $5200$~d, $260$~d, $44.4$~d,
$14.7$~d, $9.8$~d, and $0.737$~d. This is the basic period set in all combinations
revealed by FREDEC. The combination with the largest number of non-diurnal periods
contains only these six components, which offer almost the best fit (Vuong statistic of
$0.01$, maximum multifrequency $\FAP$ of $0.7\%$). Other alternative combinations involve
a subsample of this basic combination, complemented by some diurnal aliases. All of these
basic periods are orbital periods of the known five planets, except for the period of
$9.8$~d (alternatively $1.11$~d).

This additional period of $9.8$~d could represent a hint of some previously unknown planet
of the system, so we undertook a more detailed investigation of this variation. Our
preliminary conclusion is that this is not necessarily a planet-induced variation. It may
represent an artifact of the multiplicative noise model, in which the weights $w_i$ are
assumed known, and the true uncertainties are assumed equal to $\kappa/w_i$ with a common
scale factor $\kappa$. For exoplanetary radial velocity fits a better noise model is the
additive one, where the error variances are equal to the sum of some known instrumental
part and of the ``jitter'' \citep{Wright05,Baluev08b}. We find that the $9.8$~d peak is
indeed present in the periodograms constructed using the classic noise model, but it
disappears completely when the additive noise model is adopted (using the method of
\citealt{Baluev08b}). Previously we noted that the additive noise model may introduce
significant changes for heterogeneous time series, in which the jitter may appear differ
for different subsets (coming e.g. from different observing teams). So far we have not yet
seen case demonstrating that the choice noise model may become so important for a single
homogeneous dataset. This however does not decrease the value of our new FREDEC algorithm,
since it is a general-purpose data-analysis tool not designed to deal with a special task.

However, we still do not close the question of the reality of the new $9.8$~d period in
these data. It is very suspicious that this period appears in a 3:2 commensurability with
the another planetary period of $14.7$~d. We could not explain the $9.8$~d variation by
applying the multi-Keplerian model with non-zero eccentricities or the Newtonian model
involving planet-planet perturbations in the system. It is relatively unusual that this
period disappears only after applying a special model to the RV noise that does not
redistribute the power across the frequencies (like e.g. the model of a correlated noise
would do). In fact, we are not aware of any work clearly and undoubtfully showing that the
additive noise model is indeed practically superior over the classic multiplicative one.
So far, the additive noise model was an priori likely but unverified assumption.
Therefore, we believe the hypothesis of the new $9.8$~d planet in the 55~Cnc system needs
a further detailed investigation.

We pay so much attention to any tiny hints of additional putative planets orbiting 55~Cnc
because of the recent attempts to fit this system to a Titius-Bode-like law
\citep{PovedaLara08}. Such hypotheses appear very endurant regardless of all controversies
and disputes around them. This is probably because they offer an apparently easy way to
predict new planets in known multi-planet systems. The existence of the $9.8$~d planet
orbiting 55~Cnc would represent a further argument against the predictive power of any
``law'' of such type. \citet{PovedaLara08} did predict new planets in this system, but at
a much larger period values like $3.1$~yr and $62$~yr, where our algorithm finds nothing.
On contrary, they did not predict anything at the period of $9.8$~d which is now the next
planetary candidate in the queue.

\section{Conclusions}
\label{sec_conc}
We believe that regardless of the limitations that we have mentioned above, the FREDEC
algorithm still might be very useful in practice. To our concern it is the only available
algorithm that meticoulosly considers the entire set of all possible frequency
combinations, e.g. including the variations that might be wrongly interpreted as aliases.
Also, it is the only algorithm that deals with complete false alarm probabilities of the
multifrequency combinations. The option of GPU parallelization might be also very helpful.
The practical usage of the FREDEC algorithm is easy, as it is entirely automatic.

We expect this software will be useful in many astronomical applications, such as search
of exoplanets in radial velocity data and investigation of variable stars. It can also be
helpful in the fields other than astronomy, that deal with the period search task, e.g.
geophysics and climatology.

\section*{Acknowledgements}
This work was supported by Russian Foundation for Basic Research (project 12-02-31119
mol\_a) and by the programme of the Presidium of Russian Academy of Sciences
``Non-stationary phenomena in the objects of the Universe''. I am grateful to the
anonymous reviewer for providing very useful suggestions concerning the manuscript.

\bibliographystyle{model2-names}
\bibliography{fredec}







\end{document}